\documentclass[reprint, superscriptaddress, amsmath,amssymb, aps]{revtex4-2}

\usepackage{graphicx}% Include figure files
\usepackage{dcolumn}% Align table columns on decimal point
\usepackage{bm}% bold math
\usepackage{chemformula}
\usepackage{amsmath,amsthm,amssymb}
\usepackage{gensymb}
\usepackage{xr}
\usepackage{hyperref}% add hypertext capabilities
\usepackage{soul}
\usepackage[mathlines]{lineno}% Enable numbering of text and display math
\begin{document}

\title{Mott materials: unsuccessful metals with a bright future}

\author{Alessandra Milloch}
\email[]{alessandra.milloch@unicatt.it}
\affiliation{Department of Mathematics and Physics, Università Cattolica del Sacro Cuore, Brescia I-25133, Italy}
\affiliation{ILAMP (Interdisciplinary Laboratories for Advanced Materials Physics), Università Cattolica del Sacro Cuore, Brescia I-25133, Italy}
\affiliation{Department of Physics and Astronomy, KU Leuven, B-3001 Leuven, Belgium}

\author{Michele Fabrizio}
\email[]{fabrizio@sissa.it}
\affiliation{Scuola Internazionale Superiore di Studi Avanzati (SISSA), Via Bonomea 265, 34136 Trieste, Italy}

\author{Claudio Giannetti}
\email[]{claudio.giannetti@unicatt.it}
\affiliation{Department of Mathematics and Physics, Università Cattolica del Sacro Cuore, Brescia I-25133, Italy}
\affiliation{ILAMP (Interdisciplinary Laboratories for Advanced Materials Physics), Università Cattolica del Sacro Cuore, Brescia I-25133, Italy}
\affiliation{CNR-INO (National Institute of Optics), via Branze 45, 25123 Brescia, Italy}

\begin{abstract}
Achieving the full understanding and control of the insulator-to-metal transition in Mott materials is key for the next generation of electronics devices, with applications ranging from ultrafast transistors, volatile and non-volatile memories and artificial neurons for neuromorphic computing. In this work, we will review the state-of-the-art knowledge of the Mott transition, with specific focus on materials of relevance for actual devices, such as vanadium and other transition metal oxides and chalcogenides. We will emphasize the current attempts in controlling the Mott switching dynamics via the application of external voltage and electromagnetic pulses and we will discuss how the recent advances in time- and space-resolved techniques are boosting the comprehension of the firing process. The nature of the voltage/light-induced Mott switching is inherently different from what is attainable by the slower variation of thermodynamic parameters, thus offering promising routes to achieving the reversible and ultrafast control of conductivity and magnetism in Mott nanodevices.
\end{abstract}
 
\maketitle

Mott insulators are unsuccessful metals with a large density of carriers that are made inactive by Coulomb repulsion but could become available for electric conduction under proper external stimuli. Moreover, the electron localisation favours the emergence of other phenomena that are hindered in metals, most notably magnetism. This richness envisions potential applications \cite{Tokura2017,Wang2020} in the realm of electronics, spintronics and sensing, with a strong drive towards developing neuromorphic devices for the hardware implementation of neural networks \cite{mehonic2022} or for ultrafast volatile and non-volatile memories or processors \cite{waser2007nanoionics,ielmini2016resistive,Ran2023}. 
Most efforts have been therefore devoted to finding easy ways to drive insulator-to-metal Mott transitions (IMMTs) that are reversible and may operate at frequencies as high as several THz, thus much faster than varying thermodynamic variables, such as pressure or temperature. 
Promising candidate materials include transition metal oxides, such as nickelates, manganites, vanadium oxides \cite{Imada1998,sawa2008resistive,WANG201963}, and chalcogenides, such as GaTa$_4$Se$_{8-x}$Te$_x$ \cite{guiot2013avalanche} and 1$T$-TaS$_2$ \cite{yoshida2015memristive}. 
The tools that have been developed to induce and investigate the IMMT range from static or pulsed bias voltages to electromagnetic radiation, ranging from THz to X-rays. 
%In these cases, the peculiarities of Mott insulators give rise to intriguing phenomena that go beyond what can be achieved simply by changing temperature or pressure. The intrinsic first order nature of the IMMT allows the existence of metastable phases that can be stabilised away from thermal equilibrium leading to non-trivial spatial and temporal profiles. \\
%The ultimate goal of the research in this field is to achieve the full and reversible control of the Mott switching and realize ultrafast electronic and spintronic devices capable of operating at frequencies as high as several THz. 
In this work we will review the recent advances in the understanding of the insulator-to-metal switching processes in Mott materials under the application of external fields, as well as the current knowledge of the transition dynamics in both real space and time domain along with  possible routes towards the full control of the Mott switching process.

\section{Theoretical framework}

\begin{figure*}[t]
\includegraphics[width = 14 cm]{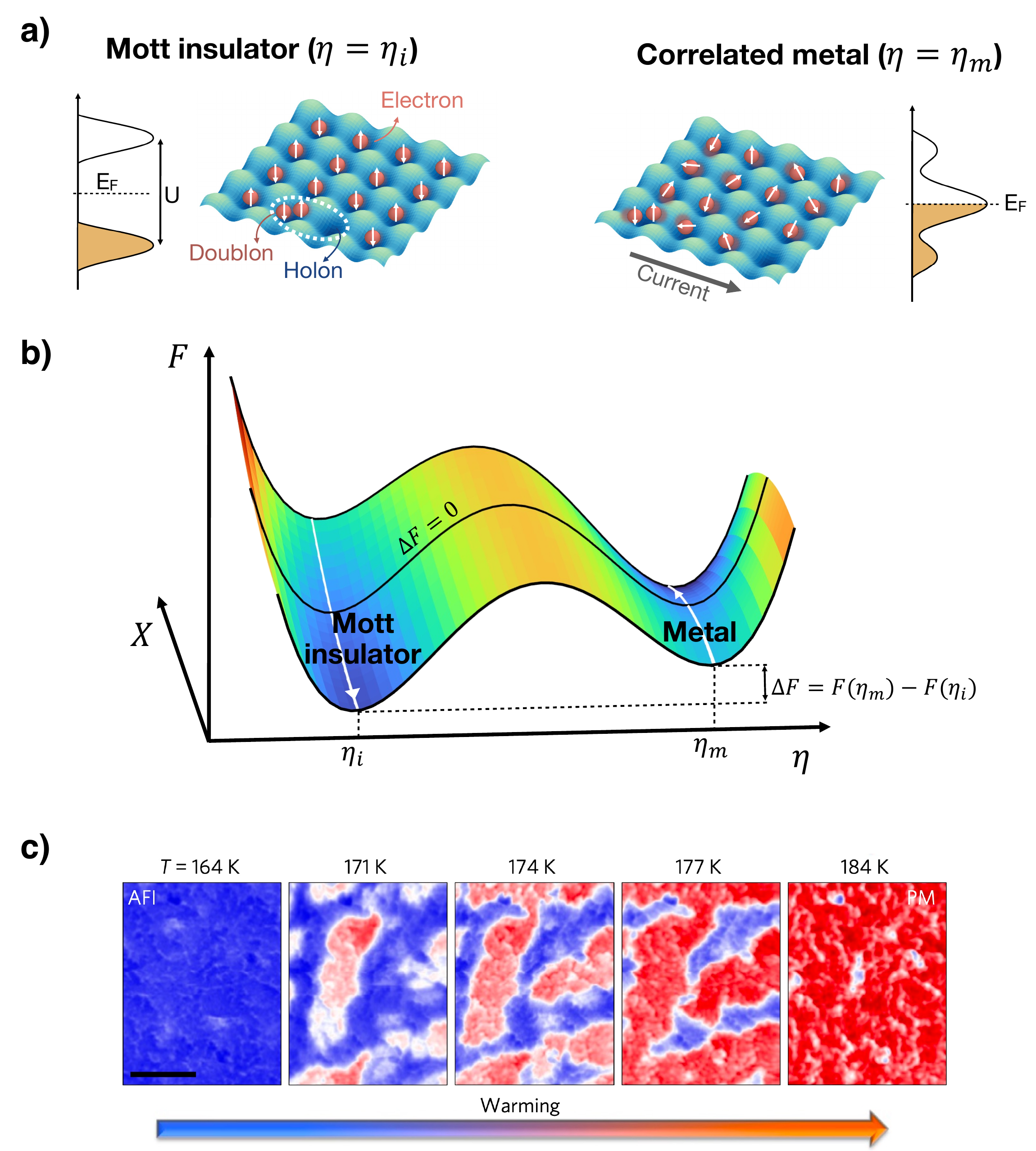}
\caption{\textbf{First-order insulator-metal transition in Mott materials.} a) Sketch of the Mott insulating state (left panel) and metallic state (right panel) in the single-band Hubbard model. In the Mott insulating phase, the single-particle spectrum comprises lower and upper Hubbard bands separated by a gap of the order of the on-site Coulomb repulsion $U$. The electrons are frozen in the lattice periodic potential (blue surface) and their spin ordered antiferromagnetically. Negatively charged doublons and positively charged holons are bound so that a weak electric field cannot drive a finite current. Upon transition into the metal phase, magnetic order is lost, a fraction of unbound doublons and holons forms a narrow quasiparticle band at the Fermi energy $E_F$ that can carry current in an electric field.  b) Free energy functional $F$ at local thermal equilibrium plotted as a function of the system variable $\eta$, which includes electronic and lattice configurations, and the control parameter $X$, representing either the applied voltage or the density of selected particle-hole excitations. As $X$ is increased, the free energy difference $\Delta F>0$ between the metallic local minimum and the insulating global one decreases until $\Delta F$ crosses zero and the metal turns into the global minimum. c) Phase coexistence in \ch{V2O3}, adapted from Ref. \citenum{McLeod2017}. Near-field infrared microscopy shows the evolution, upon heating, from the antiferromagnetic insulator (AFI, blue) to paramagnetic metal (PM, red) phase; scale bar, 1 \textmu m.}
\label{fig:energyfuncional}
\end{figure*}

The Mott transition arises from the competition between the band energy gain, maximized by conduction electron delocalization throughout the lattice, and the Coulomb repulsion that instead favours charge localization. At thermal equilibrium, the outcome of such competition can be controlled by charge doping, physical and/or chemical pressure, and by temperature, in which case the entropy contribution to the free energy becomes important and may drive by its own the metal-insulator transition. As a matter of fact, a Mott insulator is not characterised like conventional band insulators by the existence of an energy gap around the chemical potential where wave packet propagation is forbidden. Instead, any Mott insulator accommodates, even at very low temperature, a sizeable concentration of oppositely charged fluctuations, i.e., a local excess or reduction of the conduction electron density driven by quantum fluctuations \cite{Mott-1949,Sandro-VMC-PRL2005}. In the simple half-filled Hubbard model, these fluctuations correspond to doubly-occupied and empty sites, so-called \textit{doublons} and \textit{holons}, respectively (Fig. \ref{fig:energyfuncional}a).  However, positive (holon) and negative (doublon) charge fluctuations are bound together by Coulomb interaction in the Mott phase; their binding energy being just the gap between upper and lower Hubbard bands in models with short-range repulsion. Therefore, a threshold electric field is required to move them far apart and thus induce electric current. Upon approaching the Mott transition, the density of those fluctuations increases, which entails a corresponding raise of Coulomb screening until, at the transition, a fraction of holons and doublons unbind and form a narrow quasiparticle band pinned at the chemical potential \cite{DMFT-review,Daniele-exciton-PRM2019}. This is in essence the argument originally put forth by Mott \cite{Mott-1949}. 
\\   
Although one can conceive \cite{Mott-1949} and  
study \cite{Brinkman-Rice-PRB1970,DMFT-review,Sandro-VMC-PRL2005,Nicola-ghostGA-PRB2017} a hypothetical Mott transition 
without side effects, the latter are inevitable in real materials. Indeed, even in the simplest 
example of an isostructural non-magnetic metal-to-insulator Mott transition, one generally expects a volume expansion, like, e.g., in chromium doped V$_2$O$_3$ at ambient temperature \cite{McWhan1969}. However, most of the observed Mott transitions driven by temperature or pressure are accompanied by spontaneous symmetry breaking phenomena: magnetism and/or lattice distortions changing the crystal symmetry. For instance, V$_2$O$_3$ undergoes at $T_c\simeq 170~\text{K}$ a transition from a rhombohedral paramagnetic metal into an antiferromagnetic insulator with a monoclinic structure that breaks the threefold rotational symmetry \cite{Dernier1970,McWhan1970}. Vanadium dioxide, VO$_2$, displays near room temperature ($T_c\simeq 340~\text{K}$) a transition from a paramagnetic high-temperature metal to paramagnetic low-temperature insulator \cite{VO2-magn_susc-Nature1954}. This transition is accompanied by a lattice transformation from the metal rutile structure into the insulator monoclinic one \cite{VO2-NanoLett2017}, where the vanadium atoms move both along the $c$-axis, leading to a Peierls distortion, and within the $a-b$ plane, which raises the crystal field splitting between $a_{1g}$ and $e^\pi_g$ orbitals, both displacements being crucial to stabilise the correlated insulating state  \cite{Goodenough,Adriano-VO2-PRR2020}. We mention that the metal has not always a larger crystal symmetry than the insulator. For instance, at 
ambient pressure GaTa$_4$Se$_8$ is a narrow gap Mott insulator which turns into a metal with lower space group symmetry under pressure\cite{GaTa4Se8-JPCL2021}.\\
Another common feature of Mott transitions in real materials is their first order nature (Fig. \ref{fig:energyfuncional}b), which was already predicted by Mott \cite{Mott-1949} and confirmed in 
Hubbard-like models \cite{Castellani&Ranninger-PRL1979,DMFT-review,Attaccalite-GA-PRB2003,Werner&Millis-PRL2007,Federico-PRB2011}. Moreover, 
the unavoidable involvement of the lattice degrees of freedom is expected to strongly enhance the discontinuous character of the transition. For instance, the rhombohedral-monoclinic metal-insulator transition in V$_2$O$_3$, which is a strain-driven martensitic transformation \cite{martensitic-V2O3-1}, is accompanied by a six-order of magnitude jump in resistivity and can be quite destructive if the sample is not dealt with care. The first order nature of the  transition, irrespective of it being an inherent property of the Mott localisation phenomenon or mostly a consequence of the structural transformation \cite{entropy-phonon-VO2-Nature2014}, offers great opportunities for potential applications. Indeed, a first order Mott transition implies a region of metal-insulator phase coexistence that exists at equilibrium (Fig. \ref{fig:energyfuncional}c) and could be substantially widened in out-of-equilibrium conditions. \\  

Let us, for instance, imagine a stable Mott insulator coexisting with a metastable metal, and turn on a potential drop $V$ across the sample. 
In this situation, a droplet of the metastable metal might spontaneously 
nucleate since its high electric susceptibility $\chi_\text{e}$ implies a larger energy gain $\sim \chi_\text{e}\,E^2$ than 
the insulator in presence of the electric field $E$. 
That envisages the possibility of a genuine resistive switching, i.e., of a hypothetical field-induced insulator-to-metal Mott transition \cite{Marcelo-AdvMat2013,Giacomo-PRL2016} once the metal free energy crosses downwards the insulating one above a threshold $V_\text{thr}$. Moreover, even though at $V=0$ the metal phase were not even a local minimum of the free energy, it could become so upon increasing the field. However, $V>V_\text{thr}$ does not immediately trigger the resistive switching. For that, the metal needs not only to nucleate but also to grow till forming a percolating conducting path. When the lattice is not heavily involved, the growth can be modelled by a resistor network \cite{Marcelo-AdvMat2013} that accounts for the highly inhomogeneous distributions of the electric field and, because of Joule heating, also of the local temperature in presence of metal nuclei, leading, above an intrinsically stochastic threshold voltage \cite{Rocco2022}, to an avalanche effect with the formation of filamentary conducting channels \cite{delvalle2021spatiotemporal,Rocco2022}. This behaviour is ultimately not dissimilar to the   
dielectric breakdown that occurs away from phase coexistence \cite{Oka-PRB2012,Giacomo-PRB2015,Giacomo-PRL2016}, although it requires a voltage related to the free-energy difference between the stable insulator and metastable metal rather than the larger insulator charge gap. Nonetheless, in both insulating VO$_2$ and V$_2$O$_3$ subject to above-threshold voltage pulses there is evidence of an extremely slow relaxation to equilibrium, with a critical slowing down approaching the first-order transition temperature \cite{delValle2019}. This behaviour hints at an important role of the lattice degrees of freedom, consistent with earlier experiments in VO$_2$ field-effect transistors with ionic liquids \cite{Iwasa-APL2014} and therefore at the necessity of more realistic descriptions beyond the simple resistor network model  \cite{Tesler2018, Adriano-VO2-PRR2020,ronchi2022nanoscale}. \\ 

\noindent
The voltage bias is not the only way to drive a Mott transition away from thermal equilibrium. 
In reality, most Mott insulators have specific electronic excitations 
that, above a threshold density larger than the thermal equilibrium value, can transiently stabilise a metallic phase. This phenomenon is since long known in photoexcited semiconductors where an exciton gas to electron-hole liquid transition can be observed upon 
increasing light intensity \cite{Brinkman&Rice-exciton-PRB1973}. Also this transition can be   rationalised as in Mott's seminal work \cite{Mott-1949} or within the standard framework of a single-band Hubbard model \cite{Daniele-exciton-PRM2019}. 
However, in real Mott insulators the particle-hole excitations that more efficiently drive a transition into transient metal phases are generally inter-band transitions different from those  
across lower and upper Hubbard bands, and material specific \cite{VO2-photo-PRL2014,Lantz2017,Otto2019,Franceschini2023,murakami2023photoinduced}. For instance, in both VO$_2$ and V$_2$O$_3$ the relevant particle-hole excitations are those between $a_{1g}$ and $e^\pi_g$ orbitals of the crystal field split $t_{2g}$ $3d$-manifold. \\

\noindent
Summarizing, around a Mott transition there is metal-insulator coexistence, and that implies a free-energy landscape in the phase space, including both electronic and lattice degrees of freedom (collectively denoted as $\eta$ in Fig. \ref{fig:energyfuncional} a and b), which displays two local minima, one describing a Mott insulator ($\eta = \eta_i$) and the other a metal ($\eta = \eta_m$). Their relative depths define a free-energy difference $\Delta F$ that depends on the thermodynamic variables, doping, pressure and temperature, as well as on additional 
control parameters, e.g., the applied voltage $V$ and the density of selected particle-hole excitations (collectively denoted as $X$ in Fig. \ref{fig:energyfuncional}b). Whenever the condition $\Delta F$=0 is locally fulfilled, a metallic seed can nucleate and trigger a switching process that can be ultrafast and reversible in nature. Although the model here discussed directly applies to the Mott transition, it can be extended also to other systems that undergo insulator to metal transitions, such as manganese oxides, independently of the microscopic origin of the transformation.

\section{Electric-field induced Mott switching}

\begin{figure*}[]
\includegraphics[width = 14 cm]{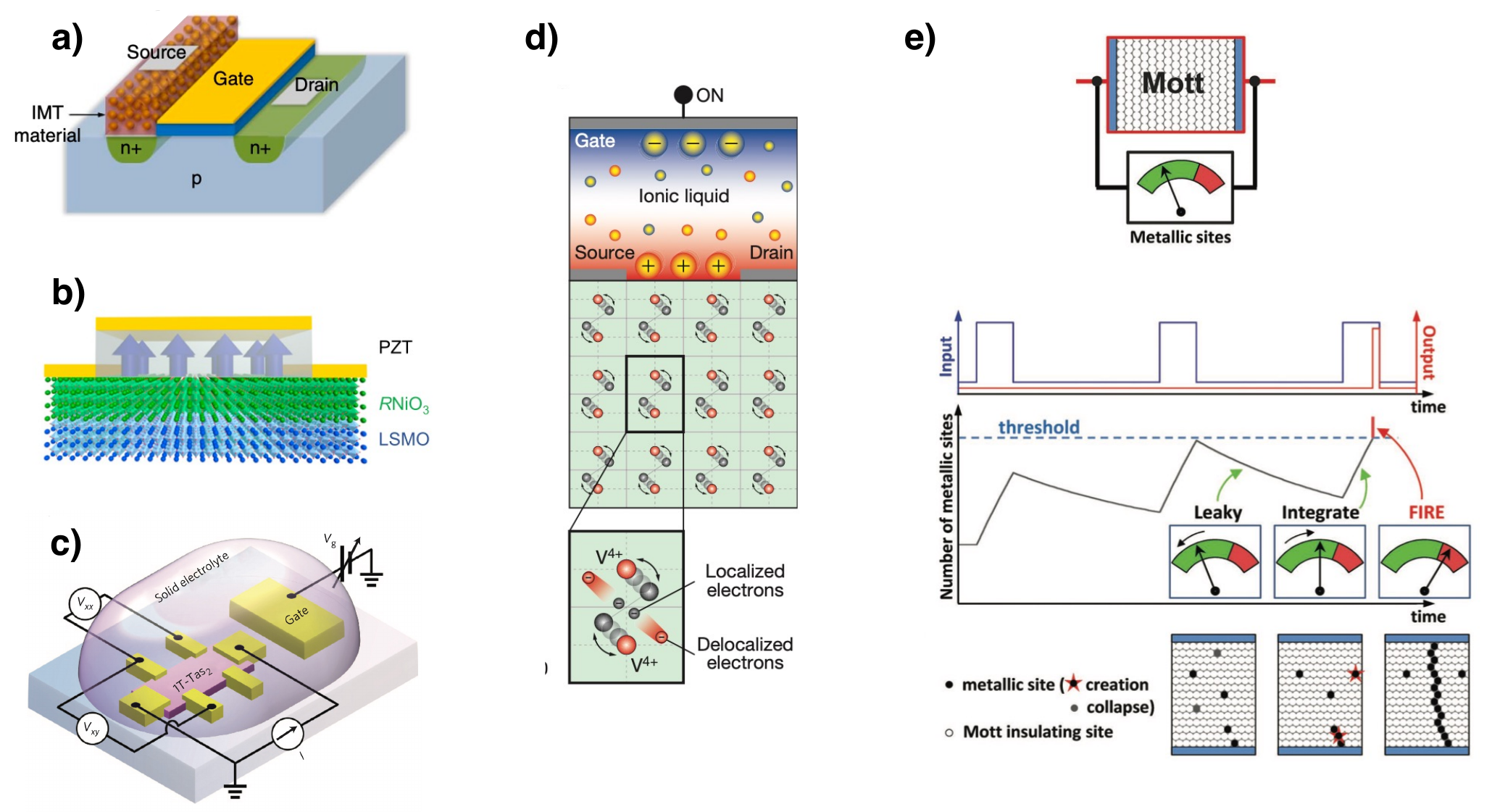}
\caption{\textbf{Mott insulator based devices.} a) Hybrid-phase-transition-FET (hyper-FET)  composed of a conventional MOSFET coupled in series with \ch{VO2}. Adapted from Ref. \citenum{Shukla2015} b) Mott ferroelectric field effect transistor (FeFET) ferroelectric composed of \ch{RNiO3}/LSMO bilayer channel (R = La, Nd, Sm, LSMO = \ch{La_{0.67}Sr_{0.33}MnO3}) gated via the ferroelectric \ch{PbZr_{0.2}Ti_{0.8}O_3} (PZT). Adapted from Ref. \citenum{Hao2023}. c) Ionic field-effect transistor (iFET) realised with layered 1$T$-\ch{TaS2}. Adapted from Ref. \citenum{Yu2015}. d) Electric-double-layer transistor (EDLT) based on \ch{VO2}. Adapted from Ref. \citenum{Nakano2012}. e) Mott insulator hardware implementation of an artificial neuron, reproducing the leak-integrate-fire process typical of biological neurons. Adapted from Ref. \citenum{stoliar2017leaky}.}
\label{fig:devices}
\end{figure*}

The electric-field induced IMMT is a volatile phenomenon that leads to a transient non-equilibrium metal phase.  Contrary to a non-volatile resistive switching, it is not accompanied by a full structural and electronic reorganization, but just by a local collapse of the Mott gap along the conducting path \cite{Zhou2015,Janod2015,delValle2018,Wang2020}. This feature triggered many efforts to realise Mott-insulator-based transistors controlled via electric gating (Fig. \ref{fig:devices}a-d). Gate-tunable devices have been realised exploiting 1$T$-\ch{TaS2} with ionic gating \cite{Yu2015} and nickelates with ferroelectric gating \cite{Hao2023}, whereas \ch{VO2} was employed to realise prototypes of electric-double-layer transistors \cite{Nakano2012} and hybrid-phase-transition field-effect transistors \cite{Shukla2015}. In the case of ionic gating experiments, intense efforts have been devoted to clarifying the competition between the purely electronic switching and ion migration effects \cite{Jeong2013,Guan2023}. Independently of the microscopic mechanism, the strong non-linearity of the electric-field-induced IMMT makes these systems ideal platforms also for developing the hardware implementation of neural networks for neuromorphic computing  (Fig. \ref{fig:devices}e)\cite{Zhou2015,delValle2018,Ran2023,stoliar2017leaky,babich2022artificial}. Mott-insulator-based devices subject to electric pulsing implement the integrate-and-fire behaviour that mimics the spiking signals of neural systems \cite{stoliar2017leaky,Tesler2018,schofield2023harnessing,yi2018biological}.

\begin{figure*}[]
\includegraphics[width = 14 cm]{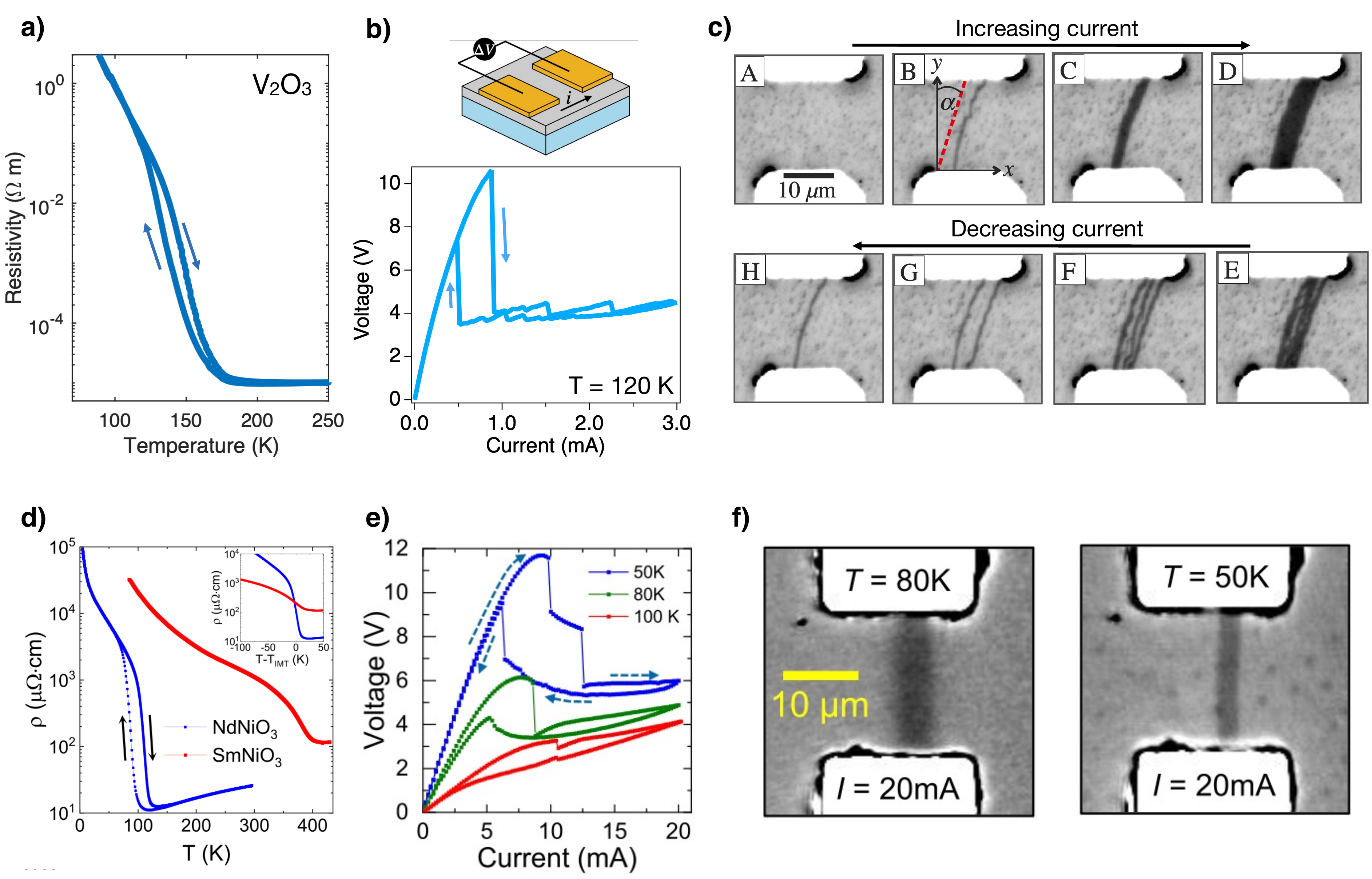}
\caption{\textbf{Imaging of filamentary resistive switching.} a) Typical resistivity hysteresis of a \ch{V2O3} film. Adapted from Ref. \citenum{milloch2024}. b) Typical current-voltage curve of a \ch{V2O3} device, sketched on the top, where resistive switching occurs at 0.9 mA upon driving of a current across a 2 \textmu m gap.  c) Wide-field optical microscopy photomicrographs of a \ch{V2O3} device acquired simultaneously to a current sweep. A  narrow metallic filament (dark line) connecting the electrodes appears above the threshold current for the switching (panel B) and widens for increasing currents (panels C-D). When the current is decreased, the metallic channel splits into multiple filaments (panels E-H) until disappearing when the current is removed and the device returns to an insulating state. Adapted from Ref. \citenum{lange2021imaging}. d) Resistivity hysteresis in nickel oxide thin films. Adapted from Ref. \citenum{luibrand2023characteristic}. e) Current-voltage curves measured in a \ch{NdNIO3} device at different temperatures. Adapted from Ref. \citenum{luibrand2023characteristic}. f) Wide-field optical microscopy images of the metallic filament forming in a \ch{NdNiO3} device upon application of an above threshold 20 mA current at temperature $T$ = 80 K (left) and $T$ = 50 K (right). Adapted from Ref. \citenum{luibrand2023characteristic}.   }
\label{fig:filaments}
\end{figure*}

\begin{figure*}[]
\includegraphics[width = 14 cm]{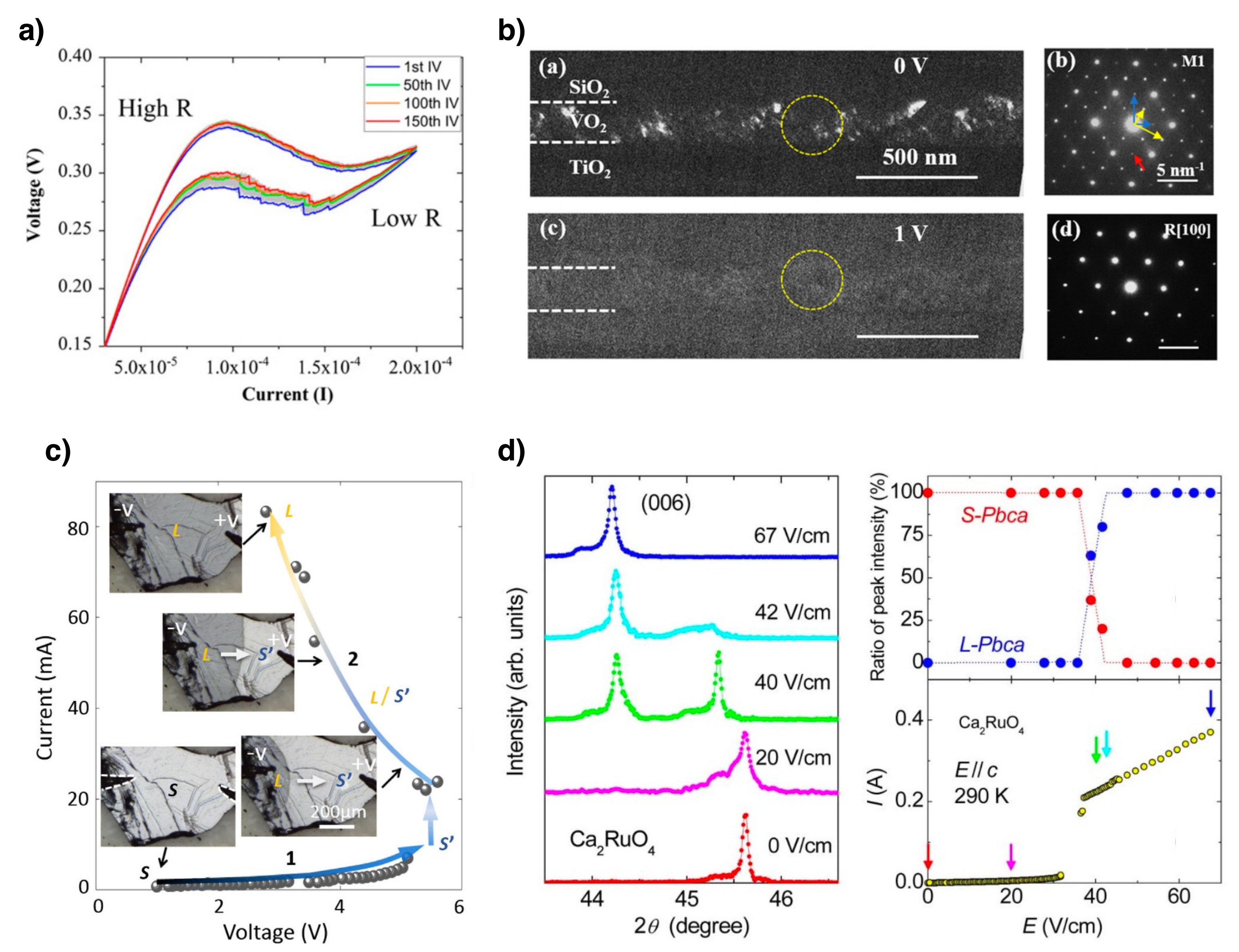}
\caption{\textbf{Structural measurements during electrical switching.} a) Voltage-current curves measured in a \ch{VO2}/\ch{TiO2} device at room temperature. Adapted from Ref. \citenum{cheng2021inherent}. b) Dark-field TEM images (left) and selected area diffraction patterns (right) of the \ch{VO2}/\ch{TiO2} device in a), measured at different values of the applied bias voltage: 0 V (top) and 1 V (bottom). Application of the above threshold bias leads to the structural transition from the monoclinic phase (M1) to the rutile one (R). The coloured arrows indicate the base vector of different monoclinic domains. Adapted from Ref. \citenum{cheng2021inherent}. c) Current-voltage curve measured on a $\sim$1 mm $\times \sim$1 mm $\times \sim$200 \textmu m \ch{Ca2RuO4} single crystal sample, along with optical micrographs showing the expansion of the metallic phase (labelled L). Adapted from Ref. \citenum{zhang2019nano}. d) Left panel: X-ray diffraction patterns showing the (006) reflections at different amplitudes of the electric field applied to a single crystal \ch{Ca2RuO4} sample. Top right panel: volume fraction of the insulting structure (S-Pbca) and metallic structure (L-Pbca) as a function of the applied electric field. Bottom right panel: current-electric field curve measured simultaneously with the X-ray diffraction. Adapted from Ref. \citenum{nakamura2013electric}.}
\label{fig:structure}
\end{figure*}

%While it has long been debated whether the resistive switching is driven by thermal (Joule heating)  or non-thermal (electric field) \cite{stoliar2014nonthermal,stoliar2013universal} mechanisms, 
The mechanism underlying the electronic resistive switching is subject of a longstanding debate. For materials undergoing a temperature-driven IMMT (see for example resistivity vs temperature plots in Figure \ref{fig:filaments}a and d), Joule heating was shown to play a key role in triggering the transition under the application of a bias voltage \cite{Mun2013,zimmers2013role,li2016joule,delvalle2021spatiotemporal,brockman2014subnanosecond}. At the same time, it was also argued that a purely thermal mechanism is not always sufficient to account for resistive switching, and electronic non-thermal mechanisms of electric-field-induced IMMT were also reported \cite{gopalakrishnan2009triggering, Marcelo-AdvMat2013, stoliar2014nonthermal, valmianski2018origin, diener2018dc, kalcheim2020non}. 
Recently, a large effort has been devoted to achieving \textit{operando} characterisation of micro and nano-devices. Pioneering microscopy experiments are opening the possibility of imaging the Mott switching process at the micro- and nano-scale in real devices. 
Optical microscopy, scanning electron microscopy (SEM), scanning microwave microscopy (SMM) and photoemission electron microscopy (PEEM) experiments, performed during the application of the electric field, allowed capturing the formation of metallic filaments short-circuiting the device in \ch{VO2} \cite{sakai2008effect,kanki2012direct,Mun2013,madan2015quantitative}, \ch{V2O3} (Figure \ref{fig:filaments} a-c) \cite{guenon2013electrical,lange2021imaging,milloch2024}, nickelates \ch{SmNiO3} and \ch{NdNiO3} (Figure \ref{fig:filaments} d-f) 
\cite{luibrand2023characteristic}. The metallic filament is observed to widen as the current flowing in the device is increased (see optical microscopy images in Fig. \ref{fig:filaments}c) and disappear once the electric field is removed and the insulating state of the system is recovered. The size of the metallic filaments depends also on the sample temperature, with filaments being wider as the critical temperature for the IMMT is approached (see Fig. \ref{fig:filaments}f) \cite{luibrand2023characteristic}. The lattice structure of the non-equilibrium state has been investigated in \ch{VO2} thin films and single crystals by means of nanoscale X-rays and electron diffraction, which revealed signals of a structural transition into the rutile high-temperature phase \cite{Mun2013,freeman2013nanoscale,cheng2021inherent}. Figure \ref{fig:structure} a-b reports structural measurements performed during application of an above threshold voltage in a \ch{VO2}/\ch{TiO2} device: the selected area diffraction patterns, obtained from transmission electron microscopy (TEM) data and shown in the left panels, reveal the structural transition from monoclinic to rutile phase upon application of a 1 V electric bias \cite{cheng2021inherent}.
One of the key parameters influencing resistive switching and filament formation is the resistivity drop across the thermally-driven IMMT, which was shown to influence the nucleation dynamics, the filament widening rate and the stochasticity of the firing process \cite{delvalle2021spatiotemporal,luibrand2023characteristic}.
Intrinsic defects also play a crucial role, pinning and controlling the formation of conductive filaments \cite{kalcheim2020non,delvalle2021spatiotemporal}. %This suggests the possibility of controlling the resistive switching by artificially creating defects in the material. 
By creating permanent defects using focused-ion-beam irradiation, it was demonstrated that it is possible to change the resistive switching mechanism from a thermal to a non-thermal field-induced IMMT \cite{kalcheim2020non}, as well as to localise the filament formation \cite{ghazikhanian2023resistive}. 
New insight into the Mott switching early stage dynamics has been provided by X-ray based microscopy techniques, sensitive to both the electronic state and the lattice symmetry breaking that accompanies the first-order Mott transition. In \ch{V2O3} thin film, PEEM imaging has shown that intrinsic topological defects in the lattice nanotexture of the threefold rotational symmetry-broken monoclinic insulator can trigger the resistive switching by pinning the filament formation and lowering the switching threshold \cite{milloch2024}. 
These results suggest that the manipulation of lattice topological defects via strain engineering can represent a new route to control the electronic switching dynamics. This idea can be extended to any system exhibiting a first-order Mott transition coupled to a symmetry-breaking order parameter \cite{milloch2024}, such as a lattice symmetry change or the onset of magnetic order.

The Mott insulator \ch{Ca2RuO4} undergoes a IMMT at 357 K or 0.5 GPa along with a structural phase transition that preserves the crystal symmetry (the metallic phase has longer lattice parameter in one direction).  In \ch{Ca2RuO4} single crystals, resistive switching takes place at very low electric field ($\sim$40 V/cm ) and is not filamentary but rather is accompanied by a bulk structural transition \cite{nakamura2013electric}. Joule heating has been ruled out as a possible mechanism for the resistive switching, which has instead been ascribed to a structural transition driven and maintained by the applied electric field \cite{nakamura2013electric,zhang2019nano,suen2023nature} and resulting into a reduction of the Mott gap \cite{suen2023nature,curcio2023current}. Nanoscale IR imaging and spectroscopy revealed also the existence of an intermediate state, maintained by dc current and characterised by a partially insulating, partially metallic phase \cite{zhang2019nano,bertinshaw2019unique}. Figure \ref{fig:structure} c-d reports electric-field induced Mott switching measurements performed in \ch{Ca2RuO4} bulk samples. Optical and IR imaging show that the metallic phase grows starting from the negative electrode and expands until the entire sample is transformed as the current is further increased (see inset images in Fig. \ref{fig:structure}c), with a characteristic striped nanotexture of alternated metal and insulating phases at the boundary between the expanding metallic region and the insulating one. X-ray diffraction data, acquired during bias application and reported in Fig. \ref{fig:structure}d,  demonstrate the occurrence of the structural transition as the applied electric field is increased above the threshold value.

\begin{figure}[]
\includegraphics[width = 8 cm]{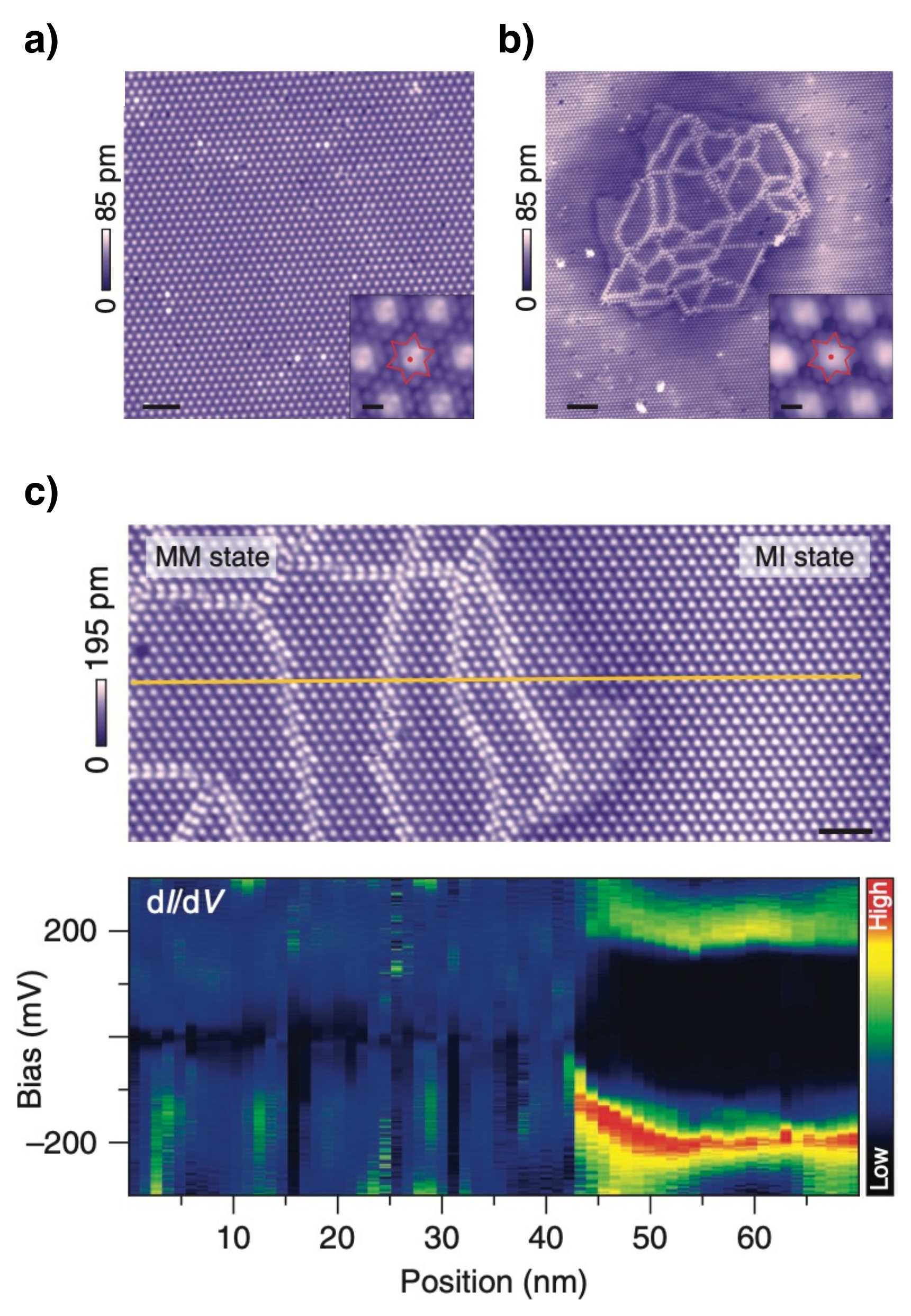}
\caption{\textbf{Voltage pulse induced metastable CDW state in 1$T$-\ch{TaS2}.} a) STM topographical image of the pristine 1$T$-\ch{TaS2} surface resolving the CCDW superlattice. Scale bar: 5 nm. The inset shows a zoomed-in view, with 0.5 nm scale bar. b) STM topography of a metastable CDW patch -  generated by a 2.8 V voltage pulse from the STM tip - in the insulating background. Scale bar: 10 nm. c) Top panel: STM image of the interface between the metastable metallic and the insulating states. Scale bar: 5 nm. Bottom panel:  Differential conductance dI/dV measured along the yellow line in the top panel, showing the insulating nature of the pristine CCDW ground state (430-meV energy gap, black), and the metallic nature of the metastable CDW mosaic state (finite local density of states at the Fermi level). Adapted from Ref. \citenum{ma2016metallic}.}
\label{fig:1T-TaS2}
\end{figure}

Chalcogenides constitute another interesting platform to unravel the complex physics of the Mott transition. 1$T$-\ch{TaS2} is a layered Mott insulator displaying various charge density wave (CDW) orders. Upon cooling, the system undergoes successive first-order phase transitions from the high-temperature metallic normal phase into more insulating phases with CDW order. An incommensurate CDW (ICCDW) sets in at 550 K, a nearly commensurate CDW (NCCDW) at 350 K and a commensurate CDW (CCDW) phase is stabilised below 180 K. In very thin 1$T$-\ch{TaS2} flakes, the NCCDW-CCDW transition is suppressed and the crystals can enter a supercooled NCCDW whose stabilisation depends on the cooling rate and sample thickness \cite{yoshida2015memristive}. Application of an in-plane bias voltage can lead to a volatile state characterised by a different charge order and lower resistance. Electrical transport measurement displaying abrupt drops in resistance signalled the possibility of resistive switching from CCDW to NCCDW \cite{yoshida2015memristive,hollander2015electrically,geremew2019bias} or to a different nearly commensurate hidden phase \cite{vaskivskyi2015controlling,svetin2017three}, from NCCDW or supercooled NCCDW to ICCDW phase transition \cite{Yu2015,zheng2017room,geremew2019bias}, as well as switching to the normal metallic state \cite{geremew2019bias}. Both Joule heating \cite{geremew2019bias,jarach2022joule} and a carrier-driven collapse of the Mott gap \cite{hollander2015electrically} have been shown to play a role in driving the resistive switching in 1$T$-\ch{TaS2}, which was reported to occur on a $\sim$3 ns timescale \cite{hollander2015electrically}. Also a non-volatile increase in resistance after the application of electric bias was reported in 1$T$-\ch{TaS2} flakes, as a result of a transition from supercooled NCCDW to CCDW and associated with the growth of CCDW domains \cite{yoshida2015memristive}. Local and reversible metallic phases can also be induced by voltage pulses from the tip of a scanning tunnelling microscope (STM) which creates a metastable CDW state over a nanometric domain, starting from CCDW \cite{ma2016metallic,cho2016nanoscale}. Atomically resolved mapping of this voltage-induced local metastable state (see Figure \ref{fig:1T-TaS2}) reveals that the new CDW phase exhibits metallic behaviour and displays a fragmented in-plane phase distribution of the CDW order parameter; the order parameter phase shift and the resulting altered stacking of CDW superlattice determine whether the new phase is insulating or metallic \cite{ma2016metallic}. 

\section{Light-controlled IMMT}

\begin{figure*}[t]
\includegraphics[width = 14 cm]{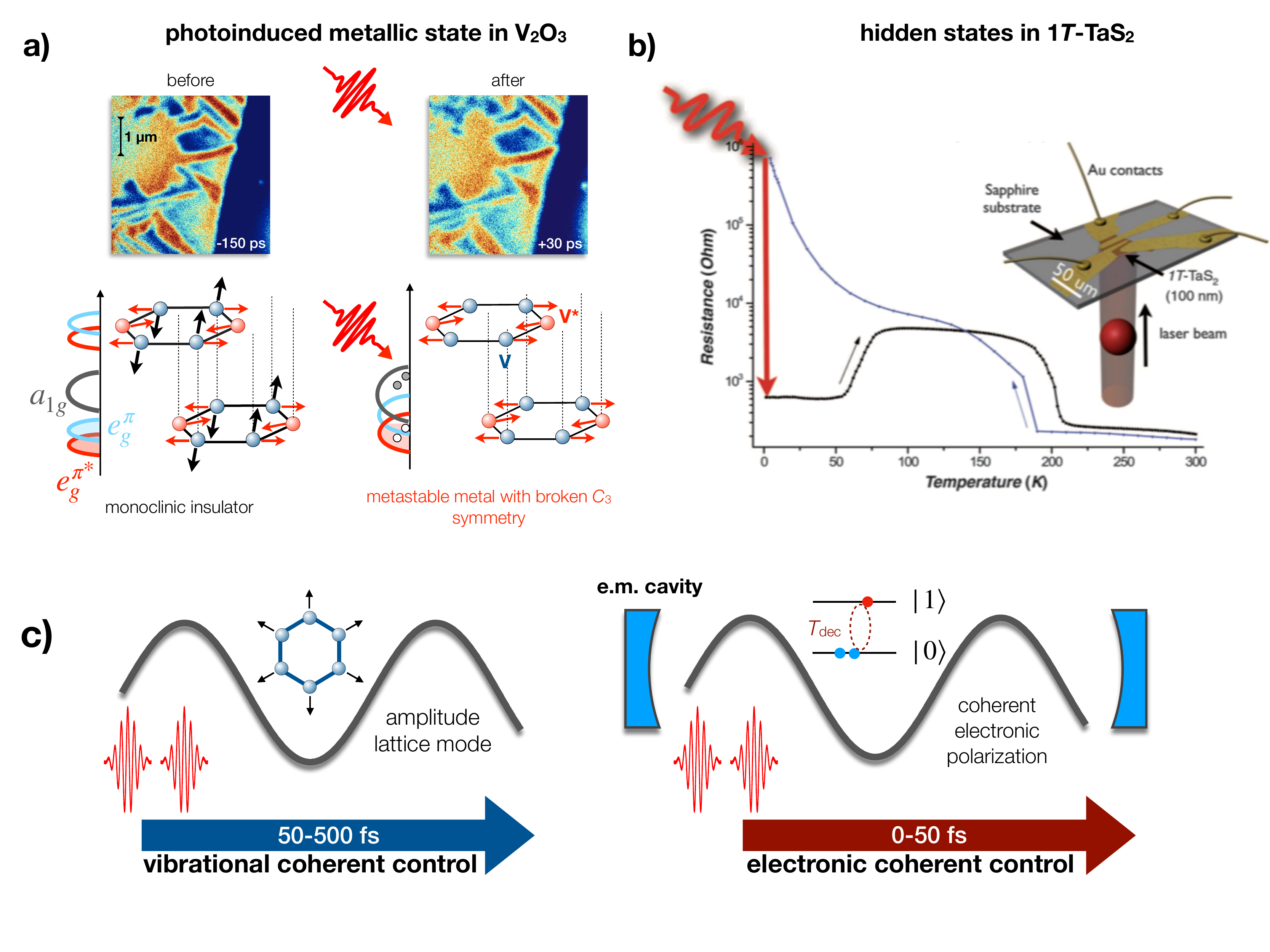}
\caption{\textbf{Photoinduced IMMT and coherent control strategies}. a) Time-resolved PEEM images of the photoinduced IMMT in V$_2$O$_3$. The top panels show the monoclinic nanotexture that spontaneously form at $T$=100 K (insulating phase). The two different domains (red and blue colors) corresponds to different in-plane monoclinic distortion along two of the three symmetry direction of the pseudo-hexagonal unit cell. After the photoinduced metallization (+30 ps) no variation of the monoclinic nanotexture is observed as compared to negative delays (-150 ps) (adapted from Refs. \citenum{Ronchi2019,ronchi2022nanoscale}). The left bottom panel illustrate a sketch of the stacked honeycomb planes in the low-temperature monoclinic insulating phase and the associated electronic bandstructure. The distortion of the hexagons(red arrows) lifts the $e^{\pi}_g$ degeneracy, whereas the out-of-plane motion of the vanadium atoms (black arrows), controls the $e^{\pi}_g$-$a_{1g}$ energy distance. The photoinduced excitation triggers the restoring of the V-V distance along the $c$-axis (tilting back of the hexagons) and consequent closing of the Mott gap, whereas it does not affect the in-plane hexagonal distortion (adapted from Ref. \citenum{Ronchi2019}). b) Drop of the resistance measured in a 1$T$-TaS$_2$-based device (inset) at 1.5 K, after excitation with a single near-infrared 35 fs light pulse. The black curve is the resistance of the photoinduced hidden conductive state, which is erased upon heating up above 60 K. The blue curve is the resistance measured on cooling. Adapted from Ref. \citenum{Stojchevska2014}. c) On the left we show a sketch of the vibrational coherent control protocol, in which a combination of multiple pulses controls the output of the IMMT during the lattice oscillation caused by a photoinduced amplitude lattice mode. The grey oscillating line represents the amplitude of the lattice mode, whose period (typically 50-500 fs) is longer than the typical time duration of light pulses. On the right we report a sketch of the faster electronic coherent control scheme, in which phase coherent electromagnetic pulses interact with a macroscopic coherent polarization state that survives on a very short timescale corresponding to the decoherence time $T_{dec}$. The coherent dynamics can be in principle modified and stabilized by means of an external resonant optical cavity. The grey line represents the temporal profile of the coherent electronic polarization, which vanishes on ultrafast timescales (0-50 fs)}.
\label{fig:timeresolved}
\end{figure*}

The possibility of exciting correlated materials with electromagnetic pulses shorter than the typical relaxation times of the electronic population has led to an intensive study of the photo-induced IMMT in the time-domain \cite{Pashkin-PRB2011,Zhang2014,Yoshida-PRB2014,Giannetti2016,Basov2017,delaTorre2021,Bao2022}. Photoinduced electronic IMMT is a common phenomenon in vanadium oxides and in low-dimensional dichalcogenides (see Refs. \citenum{Zhang2014,Giannetti2016,Basov2017,delaTorre2021} for extended reviews), which undergo abrupt electronic transformations, and in manganites \cite{Zhang2014,delaTorre2021} and ruthenates \cite{Liu2023,Rabinovich2022} where the IMMT is accompanied by an ultrafast change of the magnetic properties. For example, reversible phase switching between an antiferromagnetic charge-ordered insulator and ferromagnetic metal \cite{Fiebig1998,Takubo2005,Zhang2016,McLeod2020} can be triggered by light excitation assisted by electric field or epitaxial strain application.

Tunable excitation protocols, that can be made resonant with the electron-hole transitions associated to the Mott instability, are now mature to be combined with multiple probes sensitive to the dynamics of the electronic (e.g. optics, photoemission) and lattice/spin (e.g. electron and X-ray diffraction) degrees of freedom. The advances in time-resolved techniques are unveiling a wealth of non-thermal phenomena that challenge our understanding of the transition dynamics. Vanadium oxides are among the most interesting examples with innumerable applications in photonics and optoelectronics \cite{Basov2017}, that can be further enriched by the combination with other materials in actual devices. We cite, as an example, the recent development of the hybrid optoelectronic Mott insulator CdS/VO$_2$ \cite{Navarro2021}, the reversible optical control of IMMT across the epitaxial heterointerface of a VO$_2$/Nb:TiO$_2$ junction \cite{Yang2021} and the photo-assisted bistable switching in a two-terminal VO$_2$ device \cite{Seo2012}. Excitation of V$_2$O$_3$ in the insulating monoclinic phase with ultrafast THz pulses leads to the purely electronic metallization \cite{Giorgianni2019} within less than 1 picosecond, i.e., a timescale shorter than the percolative nucleation of the metallic phase \cite{Ronchi2019} (50-100 ps). In this case, the excitation with ultrashort THz pulses acts as a transient static electric field, that, in the metal-insulator coexistence region, can drive the stabilization of the formerly metastable metal phase \cite{Giacomo-PRL2016}. 
%Outside the coexistence, the electric breakdown occurs through a more conventional quantum tunneling across the Hubbard bands tilted by the transient THz field \cite{Giacomo-PRB2015,Giacomo-PRL2016}. 
Intensive efforts have been also put forward to address the nature of the IMMT induced in vanadium oxides by near-infrared and visible pulses. In this case, the optical pulses directly couple with interband transitions involving the $a_{1g}$ and $e^{\pi}_g$ orbitals. While the sudden change of the orbital occupation can stabilize a transient metal phase, the fast electron-phonon coupling eventually leads to the full thermalization and the restoring of the high-temperature lattice configuration of the metal. A possible non-thermal metal state in VO$_2$, which retains the monoclinic lattice structure typical of the low-temperature phase, has been claimed by ultrafast diffraction experiments \cite{Morrison2014}, and observed in model calculations \cite{Najera2017,Adriano-VO2-PRR2020}. Similar results were also reported on manganites, where a photoinduced structurally ordered state with no counterpart at equilibrium has been observed \cite{Ichikawa2011}. These results triggered a long debate about whether the electronic and lattice degrees of freedom can be transiently decoupled. Recent results suggested, on the one hand, the robust coupling between the electronic and structural transition in vanadium oxides \cite{Kalcheim2019,Vidas2020} and, on the other hand, the possibility of transiently decoupling the collapse of the Mott gap from the change of the lattice symmetry \cite{Laverock2014,Xu2022}. X-ray scattering measurements \cite{Wall2018} highlighted the important role of ultrafast lattice disordering in the transition dynamics of VO$_2$. Instead of a continuous change from the low temperature monoclinic to the high-temperature rutile lattice structures, the vanadium ions disorderly fluctuate around the equilibrium position leading to the melting of the monoclinic atomic coordinates. All these apparently contradicting results can be consequence of the inherent strong spatial inhomogeneity of the IMMT, see, e.g.,  \cite{Qazilbash2007,Qazilbash2011} in the case of VO$_2$, which initially consists in the local nucleation of metallic seeds within the insulating template. This leads to a nanotextured spatial phase coexistence \cite{Lupi2010,McLeod2017,Ronchi2019,ronchi2022nanoscale} in which topological defects can act as firing centers \cite{milloch2024}. Unfortunately, the rich physics related to spatial inhomogeneities and phase coexistence is washed out by the time-resolved techniques described so far, which return spatially integrated information. 

The recent advent of spatially and time-resolved probes is shedding new light on the nature of the early stage firing process. Time-resolved resonant X-ray microscopy was recently used to investigate the dynamics of the in-plane nanotexture formed by domains with different monoclinic distortions in V$_2$O$_3$ crystals, as shown in Fig. \ref{fig:timeresolved}a. Complementary time-resolved optical spectroscopy allowed to determine the threshold above which the IMMT is induced. Surprisingly, even though the crystal is excited with above threshold light pulses, the local monoclinic nanotexture is maintained (see Fig. \ref{fig:timeresolved}a) during the IMMT dynamics \cite{ronchi2022nanoscale}.
This result points towards a non-thermal scenario in which the light excitation melts the $c$-axis dimerization without affecting the $a$-$b$ plane monoclinic distortion.
Ultrafast X-ray nanoscale imaging of the IMMT in VO$_2$ suggests a similar scenario in which the dynamics is characterized by a transient (0-10 ps) non-thermal orthorombic structure, which relaxes towards the high-temperature tetragonal symmetry by emission of in-plane strain waves \cite{Johnson2023}. A strong interplay between strain-induced nanotexture and photoinduced IMMT dynamics has been observed in the ruthenate Ca$_2$RuO$_4$ \cite{Verma2024}. 

Another tantalizing manifestation of the Mott physics is the existence of the so called hidden states. When the Mott insulating state coexists with charge ordering, a proper combination of strain engineering and light excitation can drive the formation of hidden metallic states that do not exist at equilibrium, unlike the above mentioned metastable metal, which becomes the stable phase beyond the first order transition. This phenomenon has been observed in La$_{2/3}$Ca$_{1/3}$MnO$_3$ \cite{Zhang2016} and 1$T$-TaS$_2$ in the CCDW phase \cite{Stojchevska2014,Sun2018,Gao2022} (see Fig. \ref{fig:timeresolved}b). The extreme photo-susceptibility of these systems is such that single optical pulses can drive the formation of the hidden state. In the case of 1$T$-TaS$_2$ the highly conductive hidden state can be erased by increasing the sample temperature or by thermal annealing controlled by a train of light pulses, which is of great interest for the development of ultrafast optical switches.

\section{Perspectives}
Although the full and reversible control of the Mott switching process is yet to come, recent efforts allowed to unveil many intriguing non-equilibrium phenomena that have roots in the simultaneous existence of metallic and insulating free-energy minima, whose relative depths can be manipulated by electric fields and ultrashort electromagnetic pulses ranging from THz to the visible. The sudden change of the electronic properties during the externally driven IMMT can be exploited to manipulate the magnetic properties of materials and devices. This can be achieved by coupling Mott materials, such as vanadium oxide films, to an overlayer of magnetic materials, e.g Co, Fe, Ni or TbFeCo alloys \cite{sass2005magnetic,ramirez2016collective,ignatova2022reversible,ma2023phase}. Another strategy is based on the inherent coupling between the magnetic and electronic degrees of freedom in the same material. Although the manipulation of magnetic properties via resistive switching has been demonstrated only in magnetic manganites \cite{salev2023voltage,salev2023magnetoresistance}, we foresee that the extension of these magneto-electric hybrid schemes to Mott homostructures or Mott-magnetic heterostructures will open new frontiers in the design and development of novel spintronics devices. Many further ingredients can be added to achieve the IMMT control, such as strain engineering \cite{Zhang2016,Homm2021,Hsu2023}, coupling to electromagnetic cavities \cite{Jarc2023} (see Fig. \ref{fig:timeresolved}c,d) and combination of optical and electrical excitations \cite{Navarro2021,Ronchi2021}. Meanwhile, space- and time-resolved microscopy techniques are striving for clarifying the mechanisms that control the firing process and the electronic, magnetic and lattice properties of the transient non-equilibrium states. And even more has to come... artificial solids made of halide perovskite nanocubes \cite{Milloch2023} and low dimensional materials \cite{Chernikov2015,Yu2019,Daniele-exciton-PRM2019} are emerging as solid-state quantum simulators to reproduce and study the IMMT dynamics in a controlled environment. At the same time, coherent control protocols are expected to provide new opportunities to reversibly manipulate the IMMT in both directions. These schemes exploit photoinduced coherent lattice amplitude modes (see Fig. \ref{fig:timeresolved}c) to control the IMMT output (vibrational coherent control), as recently demonstrated for 1$T$-TaS$_2$ in the CCDW phase \cite{Maklar2023}. Purely electronic, therefore inherently faster, coherent control has been experimentally reported in V$_2$O$_3$ \cite{Franceschini2023} by resonantly exciting $e^{\pi}_g \rightarrow a_{1g}$ interband transitions and manipulating them on a timescale faster than the electronic decoherence time of the system ($T_{dec}\simeq$5-10 fs) (see Fig. \ref{fig:timeresolved}d). Furthermore, theoretical works recently suggested the possibility of combining phase-stable infrared electric fields with sub-cycle optical pulses to control the electronic phase of correlated materials \cite{Molinero2022}, with the aim of creating transient metallic states with non-thermal magnetism \cite{Lingos2021} or insulating phases starting from the metallic ground states \cite{Valmispild2024}. The availability of more than ever sophisticated spatial and time resolved  techniques is expected to boost the development of combined strategies to achieve the full and reversible control of the IMMT in actual Mott devices at frequencies as high as several THz.

\section*{Acknowledgements}
A.M. and C.G. acknowledge financial support from MIUR through the PRIN 2017 (Prot. 20172H2SC4\_005) program and from the European Union - Next Generation EU through the MUR-PRIN2022 (Prot. 20228YCYY7) program. 
C.G. acknowledges support from Università Cattolica del Sacro Cuore through D.1,  D.2.2 and D.3.1 grants.

%\section{Competing interests}
%The authors declare no competing interests.

%\section{Author Contributions}
%A.M., M.F. and C.G. prepared and reviewed the manuscript and the related figures.

\bibliography{Refs}
\end{document}